# Distribution Market Clearing and Settlement

Sina Parhizi and Amin Khodaei
Department of Electrical and Computer Engineering
University of Denver
Denver, Colorado
sina.parhizi@du.edu, amin.khodaei@du.edu

Shay Bahramirad
Smart Grid and Technology
Commonwealth Edison
Oakbrook, Illinois
shay.bahramirad@comed.com

*Abstract*— There are various undergoing efforts by system operators to set up an electricity market at the distribution level to enable a rapid and widespread deployment of distributed energy resources (DERs) and microgrids. This paper follows the previous work of the authors in implementing the distribution market operator (DMO) concept, and focuses on investigating the clearing and settlement processes performed by the DMO. The DMO clears the market to assign the awarded power from the wholesale market to customers within its service territory based on their associated demand bids. The DMO accordingly settles the market to identify the distribution locational marginal prices (DLMPs) and calculate payments from each customer and the total payment to the system operator. Numerical simulations exhibit the merits and effectiveness of the proposed DMO clearing and settlement processes.

NOMENCLATURE

| | |
|---|---|
| $a$ | Elements of the bus-line incidence matrix. |
| $b$ | Load benefit. |
| $C_c$ | Customer payments to the DMO. |
| $C_u$ | DMO payment to the ISO. |
| $C_\Delta$ | DMO cost surplus. |
| $f$ | Superscript for fixed loads. |
| $g$ | Index for bid segments. |
| $m$ | Index for distribution network buses. |
| $D$ | Load demand. |
| $P^M$ | Total assigned power from the main grid. |
| $PL$ | Line power flow. |
| $PL^{max}$ | Line power flow limit. |
| $DX$ | The amount of load awarded to each segment of the bid. |
| $PX^{max}$ | Maximum capacity of each bid segment. |
| $r$ | Superscript for responsive loads (proactive customers). |
| $t$ | Index for hours. |
| $\lambda$ | Transmission locational marginal price (T-LMP). |
| $\lambda^D$ | Distribution locational marginal price (D-LMP). |

## I. INTRODUCTION

THE GROWING proliferation of proactive electricity customers in distribution networks, which respond to system behavior by utilizing distributed energy resources (DERs) or demand response (DR), provides significant benefits for the power system while at the same time challenging the system economic and reliable operation [1][2]. On one hand, proactive customers can potentially help in shaving the peak hours, reducing transmission and distribution networks congestion levels, defer capacity upgrades, increase reliability by offering a local supply of electricity, and help the system in reaching economic objectives and meeting environmental mandates [3][4]. These benefits emphasize on the critical role of proactive customers in modern power systems. On the other hand, the objective of proactive customers is primarily to reduce their electricity payments, i.e., to increase savings, hence they tend to rely on price-based schemes for managing local generation and load resources. In simple terms, when the electricity price in the utility grid is low the customer would purchase power from the utility grid, whereas when the electricity price is high the customer would prefer to generate locally, or employ DR, and sell excess generation back to the utility grid, or reduce the local load, to increase its economic benefits. As price varies during the day, the proactive customers' load profiles will change to ensure minimum cost and maximum savings. Although highly beneficial for customers, these changes will result in an increased level of load uncertainty in the system [5], which would accordingly challenge a reliable supply-demand balance, increase day-ahead load variations and the need for load following and frequency regulation services, and result in a sub-optimal resource scheduling solution obtained by the system operator. These issues will be more noticeable as the number of proactive customers increases in the system. An additional issue that should be taken into account is the emergence of microgrids in distribution networks [6]. Microgrids, as groups of interconnected loads and DERs with clearly defined electrical boundaries and the capability to operate in the grid-connected and islanded modes, would significantly facilitate the integration of DERs as well as the engagement of DR options in distribution networks by offering a coordinate operation and control [7]–[13]. Microgrids growing deployment in power systems, i.e., an estimated capacity growth from 1.1 GW in 2012 to 4.7 GW in 2017 [7], can definitely exacerbate the concerns stemmed from a large penetration of proactive customers.

One emerging, and anticipated to be viable, solution to the mentioned problems is to manage proactive customers by introducing distribution markets that offer the capability to resolve the mentioned challenges by reducing net load variability and uncertainty [11]. A few distribution market models are currently under investigation in the United States, including the Distributed System Platform Provider (DSPP) introduced in New York via the Reforming the Energy Vision program [14] and Distribution System Operator (DSO) in California [15], [16]. Additional studies, although limited, are available in the literature including the Distribution Network Operator (DNO) model proposed in [17], the DSO model in [18], and independent distribution system operator (IDSO) in [19]. While different terminology is used to define this intermediate entity between the ISO and proactive customers, all the proposals share some common characteristics.

The distribution market operator (DMO) is an entity proposed by authors in [11][20] to facilitate the establishment of market mechanisms in distribution systems. The DMO, as an intermediate entity, communicates with the ISO and proactive customers to enable participation of customers in the wholesale market. The DMO receives the demand bids from customers in the distribution system, aggregates them and submits a single aggregated bid to the ISO. After market clearing by the ISO, the DMO divides the assigned power awarded to it between the participated customers. The DMO can be part of the electric utility company or be formed as a separate entity. In either case, it should be an independent operator so as to guarantee the fairness in market operation. The market mechanism is illustrated in Fig. 1 in which the financial transactions are performed by the DMO while the physical transactions (the actual flow of power from generation companies to loads) is performed by the load serving entity (LSE).

The implementation of a distribution market and establishment of the DMO offers several advantages for customers and the system as a whole: (i) Under distribution markets, the microgrid demand is set by the DMO and known with certainty on a day-ahead basis, which would enable an efficient control of the peak demand, increase operational reliability, and improve efficiency; (ii) The microgrid can act as a player in the electricity market and exchange power with the main grid and other customers. The DMO would facilitate microgrids market participation and coordinate the microgrids interactions with the main grid to minimize the associated operational risks and uncertainties of microgrids; (iii) There will be a considerable reduction in the required communication infrastructure in the system as the microgrids and the ISO only need to communicate with the DMOs. One important issue that needs to be taken into account is that the DMO can be formed as a new entity or be part of the currently existing electric utilities. An independent DMO would be able to set up a universal market environment instead of one for each utility. It would also be less suspected of exercising market power. On the other hand, a utility-affiliated DMO would be able to perform several functionalities currently possessed by electric utilities without necessitating additional investments. Considering the listed advantages, and many more that will be obtained using more detailed numerical simulations in future work, distribution markets can be considered as both beneficial and necessary components in modern power grids which will help accommodate a large penetration of active customers. Therefore, identifying the detailed operation of these new entities, along with efficient modeling of market clearing and settlement, is of ultimate importance, as focused on in this paper.

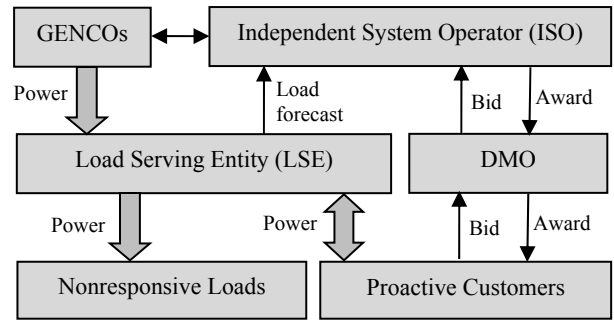

Fig. 1 Market mechanism in presence of DMO

The rest of the paper is organized as follows. The formulation for the proposed DMO market clearing is presented in Section II. The price-based market clearing is explained in section III. Settlement is demonstrated in section IV. Numerical results are presented in section V. The paper is concluded in section VI.

II. DISTRIBUTION MARKET CLEARING

Proactive customers in the distribution system offer their demand bid to the DMO. A typical bid consisting of three segments is depicted in Fig. 2. The DMO combines the individual bids and sends an aggregated bid to the ISO to be considered in the wholesale market. Once the ISO receives load bids (from DMOs) and generation bids (from GENCOs) it solves the security-constrained unit commitment and economic dispatch problems to determine the optimal unit and load schedules as well as locational marginal prices in the transmission level (T-LMP). The obtained schedule and prices are accordingly announced to DMOs and GENCOs. Each DMO would need to divide this assigned power among proactive customers in its service territory, i.e., it would "clear" the distribution market.

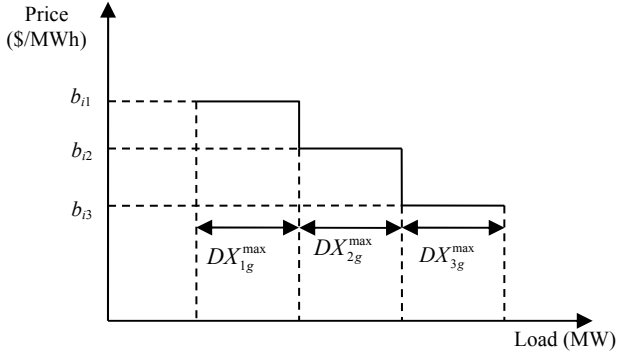

Fig. 2 Demand bid curve for customer at bus *i* with a three-segment bid

In doing so, the DMO seeks to maximize the distribution network social welfare, i.e., load benefits minus generation cost (which is the cost of assigned power from the main grid), as proposed in (1).

$$\max \sum_t \sum_i \sum_g b_{ig} DX_{mgt} - \sum_t \lambda_t P_t^M \qquad (1)$$

This objective is subject to distribution network and proactive customers prevailing constraints (2)-(6):

$$\sum_l a_{lm} PL_{lt} + D_{mt}^f + D_{mt}^r = 0 \quad \forall t, \forall m \qquad (2)$$

$$\sum_l a_{l0} PL_{lt} - P_t^M = 0 \qquad \forall t \qquad (3)$$

$$D_{mt}^r = \sum_g DX_{mgt} \qquad \forall t, \forall m \qquad (4)$$

$$0 \leq DX_{mgt} \leq DX_{mg}^{\max} \qquad \forall t, \forall m, \forall g \qquad (5)$$

$$-PL_l^{\max} \leq PL_{lt} \leq PL_l^{\max} \qquad \forall t, \forall l \qquad (6)$$

The power balance at each bus is ensured by (2) in which the power injected to each bus from connected lines is equal to the summation of loads from passive and proactive customers. The power balance at the bus connecting the distribution network to the upstream transmission network is ensured by (3) in which the power transferred by the main grid is distributed among lines connected to this bus (here the bus number 0). The load of passive customers will be constant, while that of proactive customers is variable and defined by (4). The scheduled load of proactive customers is determined based on the scheduled power consumption in each bud segment, in which each segment is limited by its associated maximum capacity (5). Line power flows are limited by the line capacity limits (6). Since the distribution network is considered to be radial, a viable distribution power flow can be guaranteed by simultaneously considering (2), (4), and (5).

The proposed market clearing model can be solved based on two completely different assumption which are based on the market design: constant power and variable power, as discussed in the following:

### A. Constant Power Clearing

The quantity of the power assigned by the ISO to the DMO is determined via the wholesale market clearing and announced to the DMO, hence it is constant. The LMP at the distribution bus is also determined by the ISO, thus the second term in (1) is constant and can be omitted from the proposed model. In this case, the DMO would distribute the assigned power to proactive customers via the proposed market clearing model. One issue that needs to be considered here is that the T-LMP does not appear in the optimization problem, hence the calculated distribution locational marginal prices (D-LMPs) can be potentially different from the T-LMP.

### B. Variable Power Clearing

In this case the DMO will schedule distribution resources based on the T-LMP determined by the ISO, however, it can deviate from the assigned power, i.e., it can purchase variable amount of power from the main grid. Therefore in the objective function (1), the main grid power $P_t^M$ is considered a variable whose value is to be determined in the optimization problem, rather than a parameter set by the ISO and given as a fixed value to the DMO. The variable power clearing model has been discussed in the literature [21][22]. Although straight-forward and easy to implement, as it utilizes the same model as ISO uses for the wholesale market clearing, this model bring about the risk of causing new peaks in the grid. When the prices are low, all customers would tend to purchase more power than they would have purchased if the prices were lower, and this would create a shortage of power. On the other hand, if the prices are too high, the customers would tend to use their own local generation to supply their load, hence creating a potential surplus. All these possible responses result in market uncertainties, hence reducing the benefits of implementing a distribution market.

### III. DISTRIBUTION MARKET SETTLEMENT

The DMO determines the D-LMPs for each distribution bus in each operation time period as a byproduct of the market clearing process. The D-LMP in each bus is calculated as the dual variable of the power balance equation in that bus (2). Using D-LMPs the market can be settled, i.e., the payments from customers and the payments to the main grid can be determined. The payment of each customer is calculated as the D-LMP times the associated load. The total customer payments is the summation of all payments as proposed in (7), in which $D_{mt}$ includes both passive and proactive customers' loads. The payment to the utility is calculated as the T-LMP times the assigned power by the ISO (8). Since losses are ignored in the proposed market clearing model, the sum of the distribution loads will be equal to the total power assigned by the ISO (9).

$$C_c = \sum_t \sum_m \lambda_{mt}^D D_{mt} \qquad (7)$$

$$C_u = \sum_t \lambda_t P_t^M \tag{8}$$

$$P_t^M = \sum_m D_{mt} \qquad \forall t \tag{9}$$

Considering the payments, the DMO cost surplus can be calculated as the difference between the two calculated payments as in (10):

$$C_\Delta = C_c - C_u = \sum_t \sum_m (\lambda_{mt}^D - \lambda_t) D_{mt} \tag{10}$$

The obtained $C_\Delta$ can be negative, positive, or zero, based on the calculated payments. The existing DMO proposals currently lack the required mechanisms, similar to what exists for the ISOs (such as auction revenue right and financial transmission right mechanisms) [23], to ensure that the cost surplus will reach zero by efficiently distributing the surplus earning to customers. This issue can potentially be a major concern regarding the fairness of the DMOs which is worthy of further investigation.

## IV. NUMERICAL SIMULATION

The IEEE 13-bus test system [24] in used to investigate the viability and the merits of the proposed distribution market clearing and settlement processes. Fig. 3 depicts this system in which proactive customers are located at buses 2, 3, 5-7, and 10- 13.

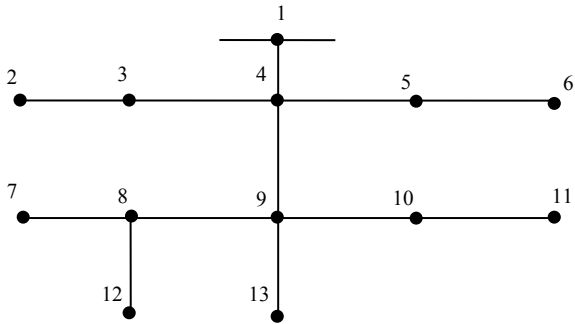

Fig. 3 IEEE 13-bus standard test system.

Three cases are considered: Case 1: Variable power market clearing and settlement; Case 2: Constant power market clearing and settlement ignoring line flow limits; and Case 3: Constant power market clearing and settlement considering line flow limits.

**Case 1:** Fig. 4 shows the hourly average D-LMP and the hourly T-LMP at the point of connection to the main grid. As shown, the D-LMPs are higher than the T-LMP in most of the hours, which is a result of the congestion in lines 5-6 and 8-7. As the prices increase, there will be a reduction in the amount of power purchased from the main grid, hence the congestion will be eliminated, and accordingly, D-LMPs and the T-LMP become equivalent. This situation happens in hours 2, 13, 14, and 16. In this case, the DMO receives $2034 from customers while paying $1900 to the ISO. The cost surplus of $134 is due to the congestion in the distribution network which needs to be distributed among customers.

**Case 2:** Fig. 5 depicts an example profile of the power assigned to the DMO via the wholesale market. The D-LMPs are determined via the distribution market clearing with constant power and based on the load bids submitted by proactive customers as $0.355/MWh. Using the given power profile, no lines are congested (resulting in equal D-LMP in all buses), the load payment is calculated as $2148, and the payment to the main grid is calculated as $2046 ($102 surplus).

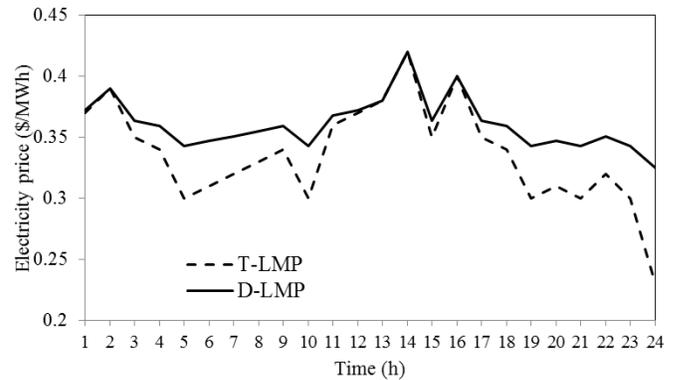

Fig. 4 LMP at the bus connected to high voltage system (T-LMP), and average of LMP at distribution system busses (D-LMPs).

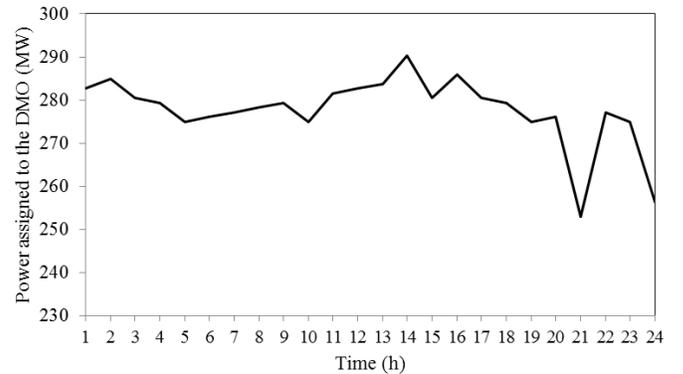

Fig. 5 The schedule of power assigned to the DMO over the 24-hour horizon.

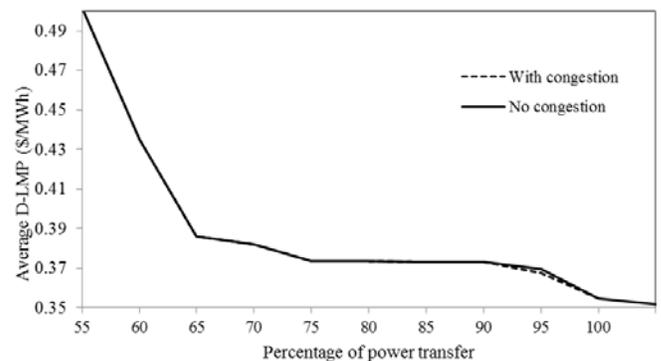

Fig. 6 Average D-LMP across the distribution network buses as a function of total assigned power.

**Case 3:** Fig. 6 shows that the average D-LMPs in the distribution network decrease as the amount of assigned power increases. Customers located at buses 10 and 13 are offering higher bids to purchase power. When line 4-9 becomes congested, the loads awarded to those customers decrease and since the bid is monotonically decreasing, as shown in Fig. 2, the prices at those buses increase. D-LMPs at buses upstream this line similarly decrease, as they will be awarded more power. Table I lists the average D-LMPs at distribution buses over the 24 hour horizon in this case. Lines 5-6 and 4-9 are congested, thus D-LMPs at buses downstream these lines (i.e., buses 6-13) have increased.

TABLE I
AVERAGE D-LMPS AT DISTRIBUTION BUSES

| Bus | 2 | 3 | 4 | 5 | 6 | 7 | 8 | 9 | 10 | 11 | 12 | 13 |
|---|---|---|---|---|---|---|---|---|---|---|---|---|
| Price | 0.278 | 0.278 | 0.278 | 0.278 | 0.353 | 0.373 | 0.373 | 0.373 | 0.373 | 0.373 | 0.373 | 0.373 |

When the power assigned to the DMO is similar to Fig. 5, the customers' payment to the DMO is $2134 and the payment to the ISO is $2046 ($88 surplus). If the assigned power is increased by 10%, the customers' payment to the DMO will be $2209, and the payment to the ISO will be $2250 ($41 surplus). This result advocates that the payment to the DMO under constant power clearing model is subject to the amount of assigned power, and independent from the T-LMP. The DMO cost surplus also can considerably change as the assigned power changes.

## V. CONCLUSION

The DMO operation was investigated in this paper in which the distribution market clearing and settlement models were proposed and formulated. Two approaches for distribution market clearing were discussed, including the constant power clearing and the variable power clearing. Although the variable power clearing model has been discussed in the literature, it was shown that this model can potentially cause variability and uncertainty in the load managed by the DMO. The constant power clearing model, on the other hand, solved this issue while bringing a potential drawback since the total payment by customers could become larger than the payment to the ISO. Both issues, however, require additional investigations as discussed and signified in this paper.